\documentclass[onecolumn,lineno,times]{bioRxiv}
\usepackage{setspace}
\let\emptyset\varnothing

\usepackage{subcaption}
\usepackage{graphicx}

\usepackage{multirow}
\usepackage{makecell}

\usepackage{comment}

\newcommand{\eq}[1]{Eq.~(\ref{eq:#1})}

\newcommand{\eqtwo}[2]{Eqs.~(\ref{eq:#1}),\ (\ref{eq:#2})}

\newcommand{\fig}[1]{Fig.~\ref{fig:#1}}
\newcommand{\figs}[2]{Figs.~\ref{fig:#1},~\ref{fig:#2}}

\newcommand{\app}[1]{appendix~\ref{app:#1}}

\newcommand{\reactto}[2][8mm]{\xrightarrow{\text{\makebox[#1]{$#2$}}}}
\newcommand{\panel}[1]{\textsf{#1}}
\newcommand{\panelb}[1]{\textbf{\panel{#1}}}

\definecolor{persianpink}{rgb}{0.97, 0.5, 0.75}

\usepackage[right]{showlabels}

\nolinenumbers

\title{The impact of simultaneous infections on phage-host ecology}

\author[1,a,b]{Jaye Sudweeks} 
\author[a,b]{Christoph Hauert}

\affil[1]{Corresponding author. email: sudweeks@math.ubc.ca; postal address: 1984 Mathematics Rd, Vancouver, BC, V6T 1Z2, Canada}
\affil[a]{Department of Mathematics, University of British Columbia, 1984 Mathematics Road, Vancouver B.C. Canada, V6T 1Z2}
\affil[b]{Department of Zoology, University of British Columbia, 6270 University Boulevard, Vancouver B.C. Canada, V6T 1Z4}


\begin{abstract}
\nolinenumbers
 Phages use bacterial host resources to replicate, intrinsically linking phage and host survival. To understand phage dynamics, it is essential to understand phage-host ecology. A key step in this ecology is infection of bacterial hosts. Previous work has explored single and multiple, sequential infections. Here we focus on the theory of simultaneous infections, where multiple phages simultaneously attach to and infect one bacterial host cell. Simultaneous infections are a relevant infection dynamic to consider, especially at high phage densities when many phages attach to a single host cell in a short time window.  For high bacterial growth rates, simultaneous infection can result in bi-stability: depending on initial conditions phages go extinct or co-exist with hosts, either at stable densities or through periodic oscillations of a stable limit cycle. This bears important consequences for phage applications such as phage therapy: phages can \emph{persist} even though they cannot \emph{invade}. Consequently, through spikes in phage densities it is possible to infect a bacterial population even when the phage basic reproductive number is less than one. In the regime of stable limit cycles, if timed right, only small densities of phage may be necessary.
\end{abstract}

\begin{document}

\hypersetup{pageanchor=false}
\maketitle
\thispagestyle{firststyle}
\hypersetup{pageanchor=true}

\section{Introduction}
Viruses are parasites that inject their genetic material into a host cell and hijack the host's genetic apparatus to produce copies of themselves. In particular, phages are viruses that infect bacteria. Phages are ubiquitous in nature \citep{wasik:ARM:2013} and have a widespread impact: for example, phages increase bacterial diversity \citep{abedon:CUP:2008} through
\begin{enumerate*}[label=\emph{(\roman*)}]
\item phage mediated predation: bacteria that are better competitors for resources tend to have weaker defences against phages, thus phages help to level the playing field \citep{thingstad:AME:1997};
\item transduction: host DNA is (accidentally) packaged by the virus and transferred to other host bacteria \citep{day:AbedonCUP:2008};
\item lysogenic conversion: phage DNA is integrated into the host DNA and results in changes of the host phenotype \citep{barksdale:ARM:1974,abedon:AbedonCUP:2008}.
\end{enumerate*}
Phages also disrupt nutrient cycling in aquatic ecosystems, where phage induced lysis (i.e. killing the host cell by rupturing the cell wall) renders organic carbon unavailable to higher trophic levels 
\citep{wilhelm:BioScience:1999}. In addition, phages serve as a useful model system for testing ecological theory \citep{jessup:TEE:2004,dennehy:JEB:2012,kerr:AbedonCUP:2008}. Finally, phages have a frankly astonishing number of potential applications \citep{garcia:peerJ:2023}. In the face of rising antibiotic resistant bacterial infections, perhaps the most well known  application of phages is as an alternative antibiotic \citep{kortright:CHM:2019}. However, others have proposed the use of phages to decrease bacterial antibiotic resistance and virulence \citep{leon:FrontMicrobio:2015,chan:SciRep:2016,gordillo:NatMicrobiol:2021}, treat dysbiosis (an imbalance in the gut microbial community that is associated with disease) \citep{garcia:peerJ:2023}, and even combat non-bacterial diseases \citep{krishnan:JMolBiol:2014}. Further proposed applications for phages include industrial uses as food preservatives \citep{endersen:CurrOpinFoodSci:2020}, self replicating disinfectants \citep{song:BMCMicrobiol:2021} and biofilm inhibitors (because bacteria within biofilms are often resistant to chemical disinfectants \citep{liu:Pharmaceutics:2022}), as well as use in pest control \citep{tikhe:Mbio:2022} and even as a tool to combat global warming \citep{boadi:CanJAnimSci.:2004}.  

During the lytic infection cycle, phages use host machinery to replicate. After new phage particles are produced,  the host cell is lysed (ruptured) to release phage replicates back into the environment. Crucially, because phages replicate using host cell resources they are dependent on their bacterial hosts to reproduce, and consequently the survival of a phage population is intimately tied to the survival of its host population. Thus in order to understand phages we must understand them in the context of their hosts, and ecological interactions between phages and their hosts are critical.

Phages infect host cells by adsorbing (attaching) to receptors on the host cell wall and then delivering the genomic content into the host cytoplasm. Phages are much smaller than bacteria and each host cell presents multiple receptors that phages can bind to, so multiple phages can adsorb to a single host cell, though not all adsorptions necessarily lead to infection.  Multiple adsorptions become increasingly likely at higher phage densities \citep{turner:AbedonCUP:2008,christen:Virology:1990} and can become the dominant transmission mode at sufficiently high densities \citep{turner:Nature:1999}. If phage densities are very high, it is possible that multiple phages simultaneously adsorb to and then infect the same host cell. Here, we explore the impact of simultaneous infections on phage-host ecology. We define simultaneous infection as infections that occur within a very small time window and distinguish between simultaneous infection and previously studied forms of co-infection, where after a pause an already infected host cell is infected again. Interestingly, given sufficient time phages can prevent multiple, \emph{sequential} infections through host cell manipulations \citep{joseph:EvolApp:2009} but these mechanisms are not applicable to the small time window relevant for simultaneous infections. 

 To demonstrate the relevance of  simultaneous infections consider phage therapy, or the use of phages as an antibiotic. The basic premise of phage therapy is to use phages to lyse target bacterial populations \citep{kortright:CHM:2019}. In order for a bacterial population to be eliminated, all bacteria must be infected by at least one phage. However, due to the stochastic nature of phage adsorption, in order for all bacteria to be infected at least once high densities of phage must be added, and many bacteria will be adsorbed to multiple times \citep{abedon:bacteriophage:2016}. Further, for both efficacy and in the interest of circumventing evolutionary arms races between phages and hosts \citep{hampton:Nature:2020}, it is preferable for adsorption and infection to happen quickly relative to the handling time of the sample and bacterial replication times \citep{goodridge:AbedonCUP:2008}. This creates a potential scenario where many adsorption events happen over a short time window, and it seems likely that simultaneous infection events would occur.   

Take as another example the use of phages for biodetection, or to detect their bacterial hosts \citep{goodridge:AbedonCUP:2008}. Detection can be accomplished through several methods, for example by engineering phages to carry a reporter gene, labeling phages with fluorescent dye, or monitoring for phage amplification (or an increase in phage density which indicates replication has occurred), but in all cases the detection works only if phages adsorb to target bacteria. Beyond choosing phages that are well suited for the target bacteria, adding high density of phage is a reliable way to ensure adsorption. Adding an excess of phage has the added benefit of decreasing the adsorption time. As before, this would result in many adsorptions happening in a small time window, hence increasing the likelihood of simultaneous infections.

In both the phage therapy and biodetection  examples, the crucial requirements are that bacteria are the limiting agent and that adsorption happens quickly \citep{goodridge:AbedonCUP:2008}.  Both can be accomplished by high phage densities, creating potential scenarios where many adsorption events occur in a small time window and opening the door for simultaneous infections. Further, because  phages can outnumber bacteria by an order of magnitude in many environments \citep{wasik:ARM:2013}, simultaneous infections may well be relevant in natural settings as well. If phages are concentrated around their bacterial hosts, say after a lysing event, then many adsorptions could occur in a short time frame, again creating the potential for simultaneous infections.

Here, we explore the dynamical impact of simultaneous infections on phage-host ecology. The interactions between phages and their hosts can be viewed from an ecological perspective as a predator-prey system with phages preying on bacteria, or from an epidemiological perspective with phages spreading through the bacterial population. Both perspectives provide helpful intuition and references for phage-host dynamics. In particular, for predator-prey systems we can rely on classical dynamical systems analysis, while from epidemiology we know the phages can invade when the phage basic reproductive number, $R_0$, the expected number of new phages produced by a given phage at the initial spread of infection, exceeds one \citep{nowak:Science:1996}.  Previous work in mathematical ecology has untangled some of the interwoven dynamics of phages and hosts with and without co-infection \citep{beretta:MB:1998,levin:AmNat:1977,alizon:RSIF:2013}.  Here we focus on the changes in the population dynamics arising from multiple, simultaneous infections of hosts. We use the well studied RNA phage $\varphi_6$ \citep{turner:Nature:1999} as both reference and inspiration for the model we develop and analyze.

\section{Model}
\label{section:Model} 
In the absence of phages, uninfected bacteria, $H_U$, divide at rate $b_U$,  perish at rate $d_U$, and compete for resources at rate $\xi$. This results in classical logistic growth dynamics with a net per-capita growth rate of $r= b_U - d_U$, which admits two equilibria: the trivial equilibrium $H_0^* = 0$ and carrying capacity $H_U^*=r/\xi$. 

Free living phages, $P$, can infect hosts. Single phages infect uninfected hosts at rate $\mu_1$ while two phages simultaneously infect an uninfected host at rate $\mu_2$. An alternative derivation of our model that considers only sequential infections and assumes that a second infection can only occur in a small time window after the first infection is shown in  \app{susWin}. Empirical work finds that infections are limited to few $\varphi_6$ phages per cell \citep{turner:JV:1999}. We set the limit to coinfection at two phages per host cell for mathematical convenience. Besides, very high phage densities are required for simultaneous triple infections and beyond to significantly affect the dynamics. 

After infection, phages replicate inside the host cell.  Once the phage particle count inside the host cell has reached a threshold, say $\lambda$ on average, the host lyses and releases phages back into the environment. We assume that lysing occurs once host resources are exhausted, and that the replication rate of phages is high, so the lysis time and final count of phage particles released is independent of whether one or two phages infected the host cell; accordingly, both single and simultaneous infection events result in an infected host, $H_I$.

Note that single infection events happen at rate $\mu_1\,P$ while double infections happen at rate $\mu_2\,P^2$. Consequently it seems natural that each infection in a double infection event occurs at rate $\mu_1\,P$, and hence $\mu_2\,P^2 = (\mu_1\,P)^2$ such that $\mu_2 = \mu_1^2$. In theory this allows us to eliminate one parameter by setting $\mu_1=\mu$ and $\mu_2=\mu^2$, but keeping the rates separate  provides further insights into interesting dynamical details. 

We assume that infected hosts become passive in the sense that they are no longer able to reproduce or compete for resources with other hosts and perish generally at a faster rate than their uninfected counter parts, $d>d_U$. We also assume that phage densities are sufficiently higher than bacteria such that impacts of infection events on the phage pool can be neglected, see \app{phageDense}. The rate at which infected hosts lyse is $d$, while phages degrade at the rate $d\kappa$. Hence, for $\kappa>1$ phages are shorter lived than infected hosts, while for $\kappa<1$ phages outlive infected hosts.

The events described above result in the following system of dynamical equations for the densities of uninfected hosts, $H_U$, infected hosts $H_I$ and free phage particles, $P$:
\begin{subequations}
\label{eq:modelEqs}
\begin{align}
    \dot{H}_U &= r\,H_U - \xi\,H_U^2 - \mu_1\,H_U\,P - \mu_2\,H_U\,P^2 \label{eq:dotHu} \\
    \dot{H}_I &= \mu_1\,H_U\,P + \mu_2\,H_U\,P^2 - d\,H_I \label{eq:dotHi} \\
    \dot{P} &= d\,\lambda\, H_I - d\,\kappa\, P \label{eq:dotP}.
\end{align}
\end{subequations}

\noindent See \app{FullModel} for a more detailed model derivation.

\section{Results}
\label{section:AnalysisNumerics}
In order to understand the ecological dynamics of \eq{modelEqs}, we first derive the equilibria and their stabilities to characterize the long term dynamics, followed by a bifurcation analysis to capture the distinct dynamical regimes. We use the burst size $\lambda$ as a natural choice for the bifurcation parameter because it directly impacts the rate of phage production and hence their density. Standard analytical tools are used, when feasible, and complemented by numerical investigations.

\subsection{Equilibria}
All equilibria are written in the form $E_i = (H_U^*,H_I^*,P^*)$, and labeled by the index $i$, where $H_U^*$ refers to the corresponding equilibrium density of uninfected hosts, $H_I^*$ to that of infected hosts and $P^*$ to the density of phages at this equilibrium. All three components of any biologically relevant equilibrium must be non-negative.

Two equilibria we have encountered already: the trivial equilibrium $E_0 = (0,0,0)$, which marks extinction of both hosts and phages, as well as the uninfected equilibrium $E_U = (\frac{r}{\xi},0,0)$, which applies in the absence of phages or upon their extinction.

For suitable parameters interior equilibria may exist. From \eq{dotP} it follows that
\begin{align}
\label{eq:Pstar}
    P^* &= \frac{\lambda}{\kappa}H_I^*
\end{align}
must hold at any equilibrium. Similarly, a relation between $H_U^*$ and $H_I^*$ can be derived from \eq{dotHu}. Using \eq{Pstar} and after eliminating the trivial solution (dividing by $H_U^*$), we find 
\begin{align}
\label{eq:HuStar}
    H_U^* &= -\frac{\lambda^2\mu_2}{\kappa^2\xi}H_I^{*2} - \frac{\lambda\mu_1}{\kappa\xi}H_I^* + \frac{r}{\xi}
\end{align}
Equipped with \eqtwo{Pstar}{HuStar} we can express \eq{dotHi} at any interior equilibrium solely in terms of $H_I^*$. After eliminating the trivial solution as well as common constant factors, we obtain the equilibrium densities of infected hosts, $H^*_I$, as implicit solutions given by the roots of the cubic polynomial
\begin{align}
\label{eq:cubicHi}
f(H_I) = \lambda^4\mu_2^2H_I^{3} + 2\kappa\lambda^3\mu_1\mu_2H_I^{2} + \lambda^2\kappa^2 ( \mu_1^2 - \mu_2r )H_I - \kappa^3\lambda\mu_1r + d\kappa^4\xi. 
\end{align}
Each root, $H_I^*$, refers to one of up to three interior equilibria, though none may be biologically relevant. Unfortunately the roots of cubic polynomials yield unwieldy expressions but fortunately based on the discriminant and Descartes’ rule of signs, we can still analytically determine the number of biologically relevant $H_I^*$, see \app{interior}. In summary, we obtain two dynamical scenarios based on the rate of reproduction of uninfected hosts, $r$. Each scenario results in up to two equilibrium densities of infected host cells with $H_I^*>0$: 
\begin{enumerate}[label=\emph{\Alph*.}]
\itemsep0pt
\item low bacterial growth rates, no bi-stability: $r\leq\mu_1^2/\mu_2$
	\begin{enumerate}[label=\emph{(\roman*)}]
	\item $\lambda<\lambda_T$: no interior equilibrium,
	\item $\lambda>\lambda_T$: a single interior equilibrium;
	\end{enumerate}
\item high bacterial growth rates, bi-stability possible: $r>\mu_1^2/\mu_2$
	\begin{enumerate}[label=\emph{(\roman*)}]
	\item $\lambda<\lambda_S$: no interior equilibrium,
	\item $\lambda_S <\lambda<\lambda_T$: two interior equilibria,
	\item $\lambda>\lambda_T$: a single interior equilibrium.
	\end{enumerate}
\end{enumerate}
with $\lambda_T=d\kappa\xi/(r\mu_1)$ and $\lambda_S=-d\xi\kappa\Big( 2\mu_1^3 + 9\mu_1\mu_2 r - 2(\mu_1^2 + 3\mu_2r)^{3/2}\Big)/\big(\mu_2r^2 (\mu_1^2+4\mu_2r)\big)$, see \app{interior}. Note that the biological relevance of a given $H_I^*$ does  not inherently imply that the corresponding interior equilibrium $E_i$ is biologically relevant, too. While from \eq{Pstar} we know that the sign of $P^*$ is the same as that of $H_I^*$, in contrast from \eq{HuStar} we see that, in principle, the uninfected component, $H_U^*$, of an interior equilibrium can become negative for sufficiently large infected densities, $H_I^*> \kappa \Big( -\mu_1+ \sqrt{\mu_1^2+4\mu_2r}\Big)/(2\mu_2\lambda)$. However, in \app{boundHI} we show that $H_I^*$ has an upper bound below that threshold and hence whenever $H^*_I$ is biologically relevant $H^*_U$ and $P^*$ are as well.

\subsection{Stability}
It is immediately clear that the trivial, extinction equilibrium, $E_0$, is always unstable for $r>0$ because $\dot H_U>0$ for small $P$, see \eq{dotHu}. Similarly, the uninfected equilibrium, $E_U$, is stable for $\lambda<\lambda_T=d\kappa\xi/(r\mu_1)$, which is easily confirmed by checking the eigenvalues of the Jacobian at the equilibrium, see \app{jacobian}.

A complementary and more intuitive approach to arrive at the same conclusion is to consider the basic reproductive number, $R_0$, of phages: the per capita rate of infection is $\mu_1$ and phages have an expected lifespan of $1/(d\kappa)$. Thus, a phage introduced into an entirely uninfected host cell population at its equilibrium, $E_U$ with $H_U^*=r/\xi$, infects $r\mu_1/(d\kappa\xi)$ cells on average. Note that double infections are negligible because initial phage densities are small. The expected lifespan of an infected host cell is $1/d$ and produces phages at the rate $d\lambda$. Overall, initially each phage produces $R_0=r\mu_1\lambda/(d\kappa\xi)$ new phages and phages can invade if $R_0>1$, or equivalently if $\lambda>\lambda_T$, where $\lambda_T$ the threshold for $E_U$ to become unstable.

Calculating the stability of the interior equilibria, $E_i$, using the Jacobian is analytically intractable. Instead we turn to numerical analysis for an illustrative set of parameters. For more details on parameters see \app{parameters}.

\subsection{Bifurcations}
\label{subsection:bifurcations}
The number and stability of all equilibria, except the trivial extinction equilibrium, change when varying the burst size $\lambda$, the expected number of phages produced by each infected host bacterium. Hence, $\lambda$ presents itself as an ideal and biologically relevant choice for a bifurcation parameter.

\begin{figure}[tbp]
\centerline{\includegraphics[width=0.9\textwidth]{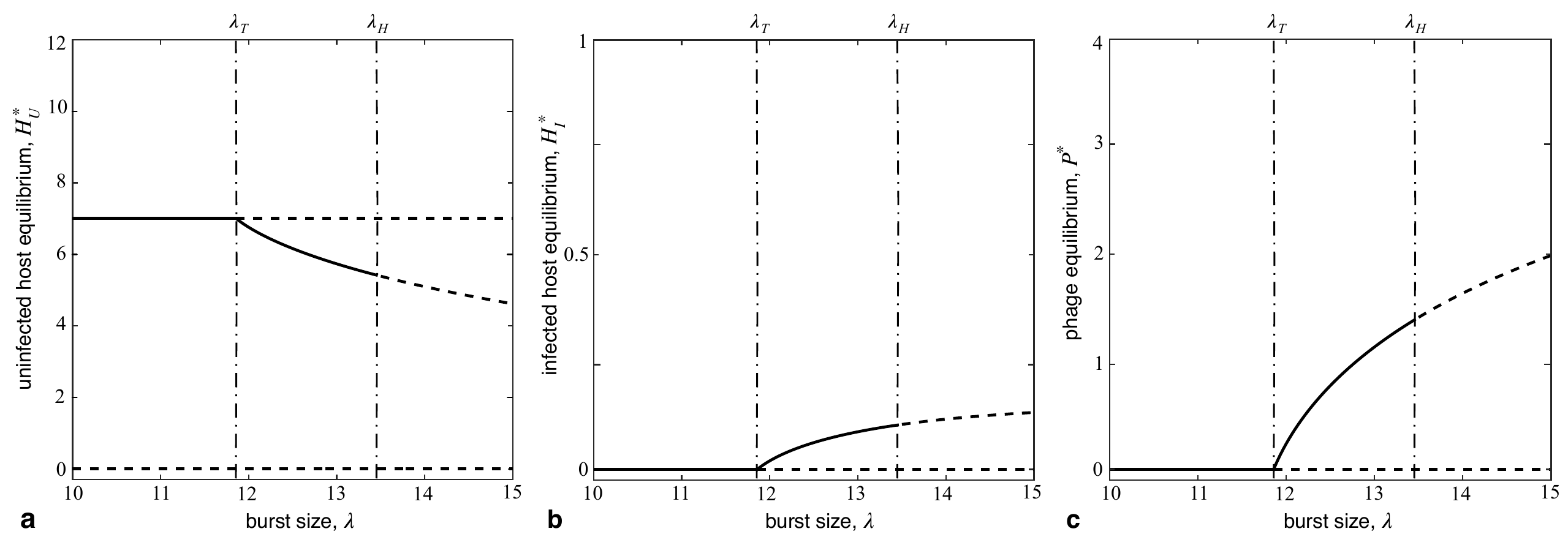}}
\caption{Bifurcation diagram for the dynamics in \eq{cubicHi} with $r<\mu_1^2/\mu_2$ as a function of the burst size $\lambda$. Each panel depicts one component of the stable (solid lines) and unstable (dashed lines) equilibria, $E_i = (H_U^*, H_I^*, P^*)$: 
\panelb{a} uninfected hosts, $H_U^*$, 
\panelb{b} infected hosts, $H_I^*$, and 
\panelb{c} phage densities, $P^*$. The trivial, extinction equilibrium, $E_0$, is always unstable and the uninfected equilibrium $E_U$ is stable for small $\lambda$. There are two bifurcations (dash-dotted vertical lines): a transcritical bifurcation at $\lambda_T=d\kappa\xi/(r\mu_1)\approx 11.86$, where a stable interior equilibrium appears and $E_U$ turns unstable, followed by a Hopf-bifurcation at $\lambda_H\approx 13.57$ where the interior equilibrium loses stability. This results in three dynamical regimes: for $\lambda<\lambda_T$ the uninfected equilibrium, $E_U$, is stable and a global attractor; for $\lambda_T<\lambda<\lambda_H$ the interior equilibrium, $E_I$, is a global attractor; and finally for $\lambda>\lambda_H$ all equilibria are unstable and stable limit cycles develop. Parameters: $r=0.7,\xi=0.1,\mu_1=0.1,\mu_2 = 0.01, d=8.3,\kappa=1$.}
\label{fig:rl1}
\end{figure}
\begin{figure}[tbp]
\centerline{\includegraphics[width=0.9\textwidth]{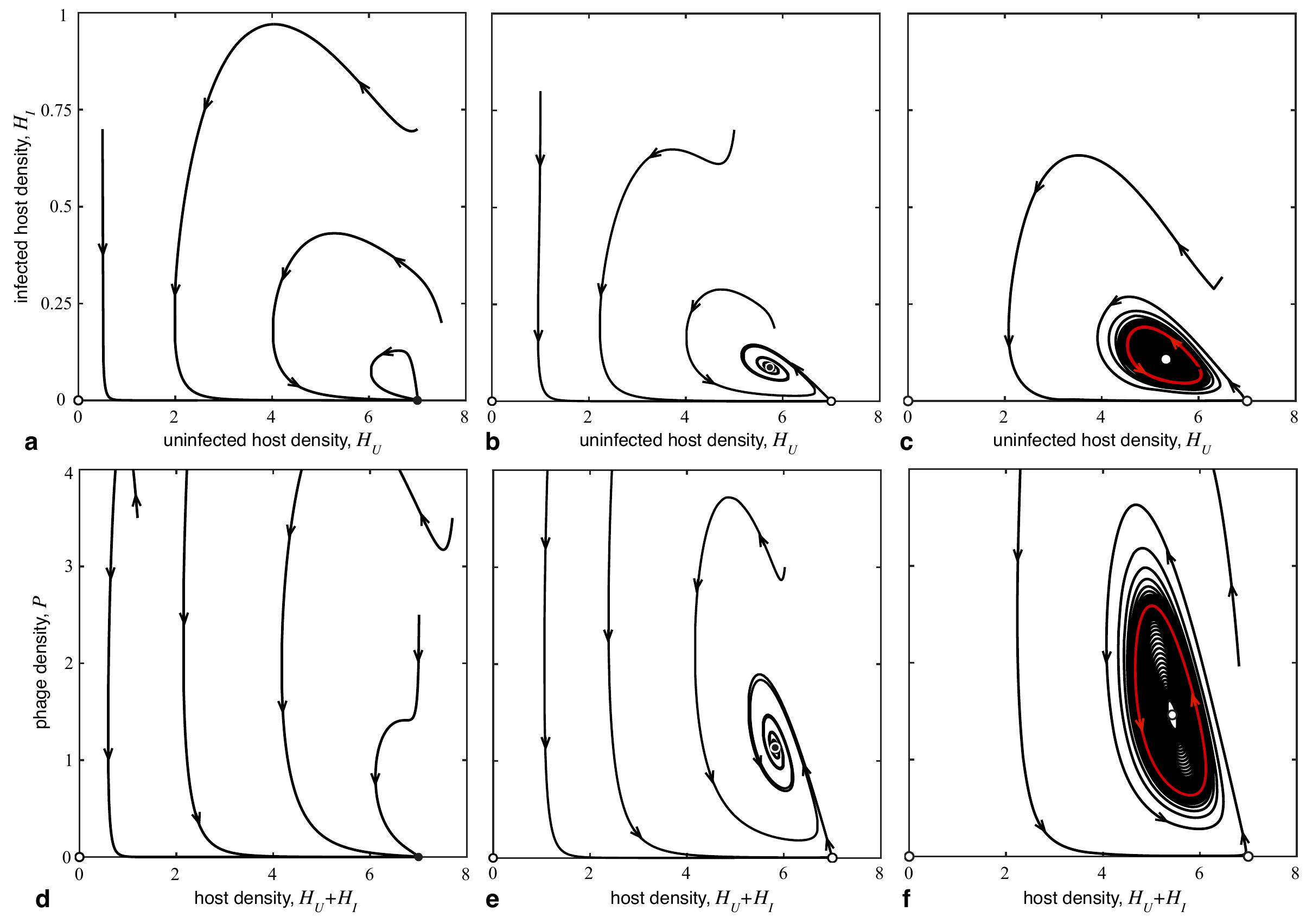}}
\caption{Projections of the phase space depicting the characteristics of the three dynamical regions for $r<\mu_1^2/\mu_2$. The top row shows infected, $H_I$, versus uninfected, $H_U$, host densities and the bottom row shows phage densities, $P$, versus host densities, $H_U+H_I$. Note that trajectories may intersect because the panels show projections of a three dimensional phase space. Dots mark stable (filled) and unstable (open) equilibria. Stable limit cycles are shown in red.
\panelb{a, d} $\lambda = 11$: all trajectories converge to $E_U$ and phages go extinct.
\panelb{b, e} $\lambda = 13$: $E_U$ becomes unstable and the stable interior equilibrium $E_I$ is a global attractor.
\panelb{c, f} $\lambda = 13.6$: the interior equilibrium $E_I$ becomes unstable and the attractor $E_I$ is replaced by a stable limit cycle.
Parameters: same as in \fig{rl1}.}
\label{fig:pprl1}
\end{figure}

 
Let us first focus on the simpler case with $r<\mu_1^2/\mu_2$, see \fig{rl1}. For small $\lambda$ the uninfected equilibrium $E_U$ is stable. At $\lambda_T$ a stable interior equilibrium $E_I$ appears through a transcritical bifurcation and the uninfected equilibrium $E_U$ loses stability. Further increases in $\lambda$ result in $E_I$ losing stability through a Hopf-bifurcation at $\lambda_H$, which is unfortunately analytically inaccessible. This yields three dynamical regimes illustrated through phase plane projections in \fig{pprl1}.

\begin{enumerate}[label=\emph{(\roman*)}]
	\item $\lambda<\lambda_T$: only the uninfected equilibrium, $E_U$, is stable. All trajectories are drawn to $E_U$, which means the phages invariably go extinct regardless of their initial density, see \fig{pprl1} \panel{a}, \panel{d}.
\item $\lambda_T<\lambda<\lambda_H$: only the interior equilibrium, $E_I$, is stable. All trajectories are drawn to $E_I$, such that host and phages stably co-exist. More specifically, because $E_U$ is now unstable, phages are able to proliferate and invade the host population, see \fig{pprl1} \panel{b}, \panel{e}.
\item $\lambda>\lambda_H$: no stable equilibrium remains. The Hopf-bifurcation together with the lack of other stable equilibria is a strong indicator of stable limit cycles, which has been confirmed numerically, see \fig{pprl1} \panel{c}, \panel{f}.
\end{enumerate}

Recall that the per-capita rate for single infections is $\mu_1\,P$ whereas for double infections it is $\mu_2\,P^2$. Hence, double infections are the dominant mode if $P>\mu_1/\mu_2$ and single infections otherwise. In \fig{rl1}c  the phage densities at equilibrium never reach the threshold for double infections to dominate. Thus in this case double infections play only a marginal role. Note that the non-generic case, $r=\mu_1^2/\mu_2$, is qualitatively the same as $r<\mu_1^2/\mu_2$.

\begin{figure}[tbp]
\centerline{\includegraphics[width=0.9\textwidth]{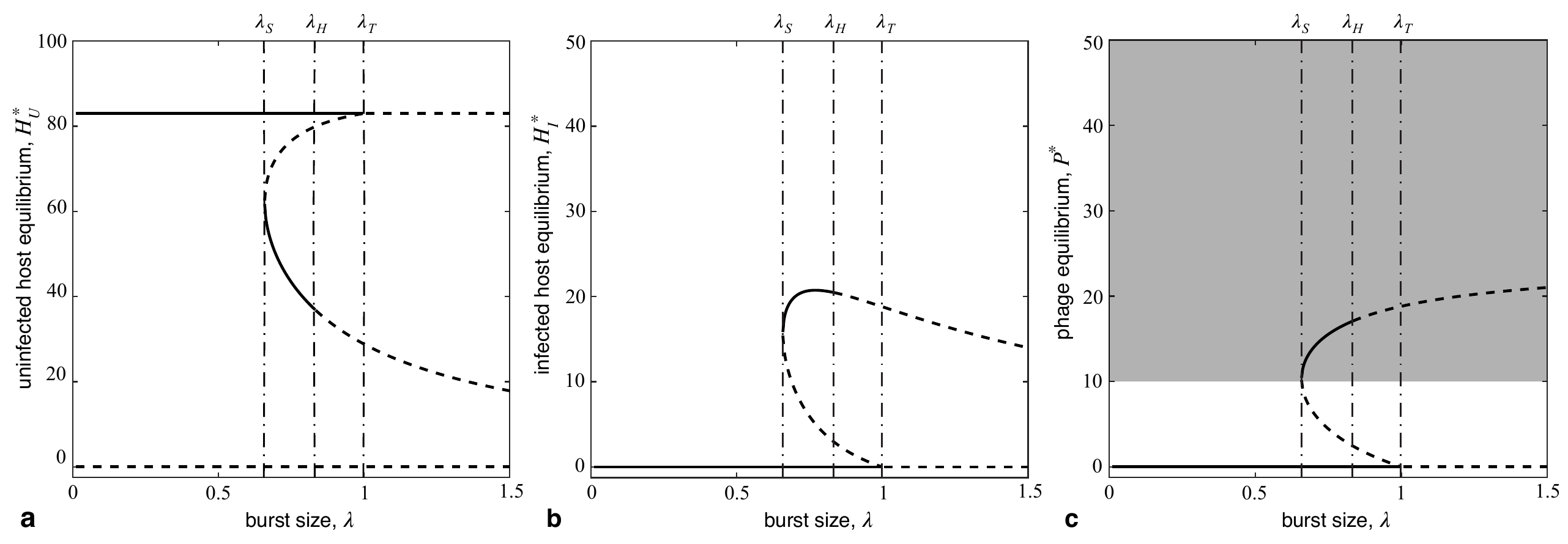}}
\caption{Bifurcation diagram for the dynamics in \eq{cubicHi} with $r>\mu_1^2/\mu_2$ as a function of the burst size $\lambda$. Each panel depicts one component of the stable (solid lines) and unstable (dashed lines) equilibria, $E_i = (H_U^*, H_I^*, P^*)$: 
\panelb{a} uninfected hosts, $H_U^*$, 
\panelb{b} infected hosts, $H_I^*$, and 
\panelb{c} phage densities, $P^*$, where the shaded area marks densities for which double infections dominate. The trivial equilibrium, $E_0$, is always unstable and the uninfected equilibrium $E_U$ is stable for small $\lambda$. There are three bifurcations (dash-dotted vertical lines): a saddle node bifurcation at $\lambda_S\approx 0.6585$ (c.f. $\lambda_-$ in \eq{lambdapm}) where a pair (stable, $E_{I^s}$ and unstable, $E_{I^u}$) of interior equilibria appear; a Hopf-bifurcation at $\lambda_H \approx 0.8329$ 
where $E_{I^s}$ loses its stability; and finally a transcritical bifurcation at $\lambda_T=1$ where the (originally) unstable interior equilibrium, $E_{I^u}$, disappears and $E_U$ loses stability. This results in four dynamical regimes: for $\lambda<\lambda_S$ the uninfected equilibrium, $E_U$, is stable and a global attractor; for $\lambda_S<\lambda<\lambda_H$ the dynamics are bi-stable with attractors $E_U$ and $E_{I^s}$; for $\lambda_H<\lambda<\lambda_T$ the dynamics remain bi-stable but the attractor $E_{I^s}$ is replaced by a stable limit cycle, see \fig{pprg1} \panel{c}, \panel{g}. Finally, for $\lambda>\lambda_T$ no stable equilibria exist but the stable limit cycle persists, see \fig{pprg1} \panel{d}, \panel{h}. Parameters: $r=8.3,\xi=0.1,\mu_1=0.1,\mu_2 = 0.01, d=8.3,\kappa=1$ (same as in \fig{rl1} except for $r$).}
\label{fig:rg1}
\end{figure}

\begin{figure}[tbp]
\centering\includegraphics[width=0.9\linewidth]{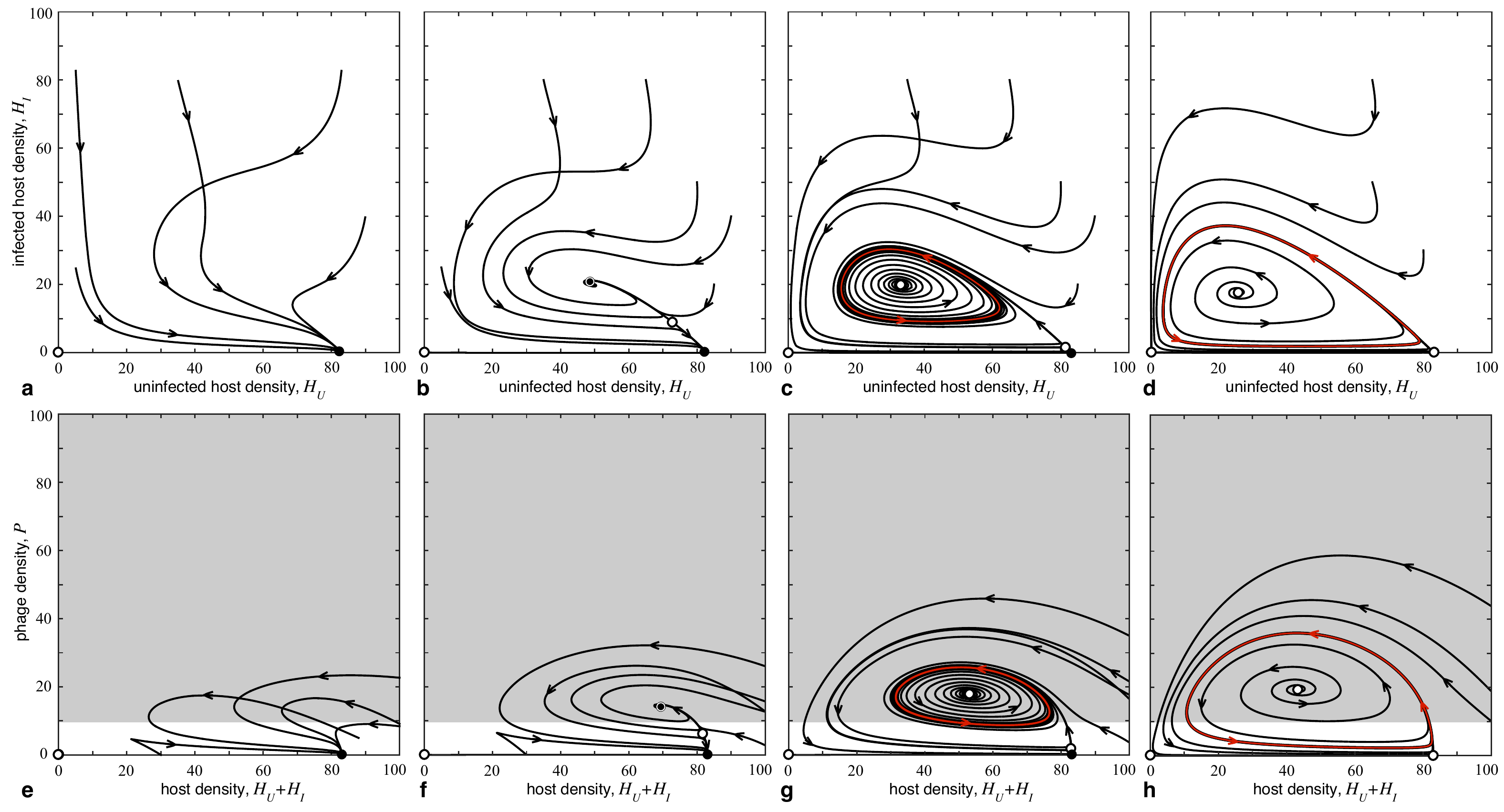}
\caption{Projections of the phase space depicting the characteristics of the four dynamical regions for $r>\mu_1^2/\mu_2$. The top row shows infected, $H_I$, versus uninfected, $H_U$, host densities and the bottom row shows phage densities, $P$, versus host densities, $H_U+H_I$. Shaded regions on the bottom row mark phage densities for which double infections dominate. Note that trajectories may cross because the panels show projections of a three dimensional phase space. Dots mark stable (filled) and unstable (open) equilibria. Stable limit cycles are shown in red.
\panelb{a, e} $\lambda = 0.5$: all trajectories converge to $E_U$ and phages go extinct.
\panelb{b, f} $\lambda = 0.7$: a pair of interior equilibria appear, $E_{I^u}$ and $E_{I^s}$, and the dynamics become bi-stable with attractors $E_U$ and $E_{I^s}$.
\panelb{c, g} $\lambda = 0.9$: the interior equilibrium $E_{I^s}$ becomes unstable and the attractor $E_{I^s}$ is replaced by a limit cycle.
\panel{d, h} $\lambda = 1.1$: the (originally) unstable interior equilibrium $E_{I^u}$ disappears and only a globally stable limit cycle remains.
Parameters: same as in \fig{rg1}.}
\label{fig:pprg1}
\end{figure}

The case with $r>\mu_1^2/\mu_2$ is more interesting because it results in richer dynamics, see \fig{rg1}. For small $\lambda$ the uninfected equilibrium $E_U$ is again stable. At $\lambda_S$ a stable/unstable pair of interior equilibria, $E_{I^s}$ and $E_{I^u}$, respectively, appear through a saddle node bifurcation. Then, at $\lambda_H$, the stable interior equilibrium, $E_{I^s}$, loses stability through a Hopf-bifurcation, leaving $E_U$ as the only stable equilibrium. Finally at $\lambda_T$ the (originally) unstable interior equilibrium, $E_{I^u}$, disappears through a transcritical bifurcation with the uninfected equilibrium, $E_U$, which loses stability at the same time. At this point no stable equilibria remain and hence stable limit cycles are expected. \fig{pprg1} depicts illustrative phase plane projections for each of the four resulting dynamical regimes.
\begin{enumerate}[label=\emph{(\roman*)}]
\itemsep0pt
\item $\lambda<\lambda_S$: only the uninfected equilibrium, $E_U$, is stable. All trajectories are drawn to $E_U$, which means the phages invariably go extinct regardless of their initial density, \fig{pprg1}\,\panel{a},\,\panel{e}.
\item $\lambda_S<\lambda<\lambda_H$: a pair of interior equilibria appears but only one, $E_{I^s}$, is stable. This results in bi-stable dynamics with two basins of attraction. Depending on the initial conditions trajectories converge to either
\begin{enumerate*}[label=\emph{(\alph*)}]
\item $E_U$, which marks the extinction of the phage population, or
\item $E_{I^s}$, indicating stable co-existence of bacteria and phages,
\end{enumerate*}
 see \fig{pprg1}\,\panel{b},\,\panel{f}.
\item $\lambda_H<\lambda<\lambda_T$: both interior equilibria are now unstable and only $E_U$ remains stable. However, this does not imply that $E_U$ is globally stable. Instead, the bi-stability remains but $E_{I^s}$ as an attractor is replaced by a stable limit cycle,  see \fig{pprg1}\,\panel{c},\,\panel{g}.
\item $\lambda>\lambda_T$: no stable equilibrium remains. The limit cycle is now a global attractor, see \fig{pprg1}\,\panel{d},\,\panel{h}.
\end{enumerate}

The major difference between the dynamics of the two cases is the emergence of bi-stability for $\lambda_S < \lambda < \lambda_T$ in the case $r>\mu_1^2/\mu_2$. In \fig{pprg1} \panel{b}, \panel{f}, for $\lambda_S < \lambda < \lambda_H$ there are two basins of attraction, and some trajectories are drawn to the stable equilibrium $E_{I^s}$, indicating stable co-existence of phage and host. Similarly, in \fig{pprg1} \panel{c}, \panel{g}, for $\lambda_H < \lambda < \lambda_T$ trajectories are drawn into a stable limit cycle, indicating fluctuating co-existence of phage and host. Crucially, note that the phage basic reproductive number $R_0 <1$ for all $\lambda_S < \lambda < \lambda_T$, meaning small densities of phage are not be able to invade and establish. However, due to bi-stability, sufficiently large densities of phage can persist and co-exist with hosts, which implies that a shock in phage densities would allow phages to get a foothold and persist.

When $r<\mu_1^2/\mu_2$ (no bi-stability) results are in agreement with previous work that considers only single infections (compare \fig{pprl1} with Figs. 1-3 of \citep{beretta:MB:1998}). This suggests that \eq{modelEqs} can be interpreted as a generalized model that encompasses this previous work while allowing investigation of the impact of simultaneous infections, see \app{parameters} for more details. The contributions of double infections lead to a major shift in dynamics at the threshold $r= \mu_1^2/\mu_2$. The threshold can be written as $(rH_U)(P^2\mu_2H_U) = (\mu_1PH_U)^2$, which admits a more intuitive interpretation in terms of two types of simultaneous events: on the left hand side replication of an uninfected host and a double infection event, and on the the right hand side  two simultaneous single infection events. Hence, we can interpret our different dynamical regimes as a balance of these simultaneous events. If the rate at which a newly emerged uninfected host is doubly infected is higher than the rate at which simultaneously two separate uninfected hosts are singly infected, then bi-stability can occur for sufficiently high $\lambda$. 

As previously noted, in \fig{rl1} \panel{c} the interior equilibrium, $E_I$, lies entirely in the region where single infections prevail. In fact, for $r< 2\mu_1/\mu_2$ single infections always dominate at $E_I$, see \app{VstarAnalysis}.  In contrast, in \fig{rg1} \panel{c} the (originally stable) interior equilibrium, $E_{I^s}$, lies in the shaded region where double infections dominate.  In general, for $r>2\mu_1^2 / \mu_2$ double infections dominate at $E_{I^s}$ for sufficiently large $\lambda$, see \app{VstarAnalysis}. Note that the transition between the two dynamical cases (no bi-stability vs. bi-stability) does not align exactly with the transition away from dominant single infections at $E_{I^s}$, and there is a region $\mu_1^2/\mu_2< r < 2\mu_1^2/\mu_2$ that exhibits bi-stability but single infections still dominate at $E_{I^s}$ regardless of $\lambda$. 

\begin{figure}[tbp]
\centering\includegraphics[height=6cm]{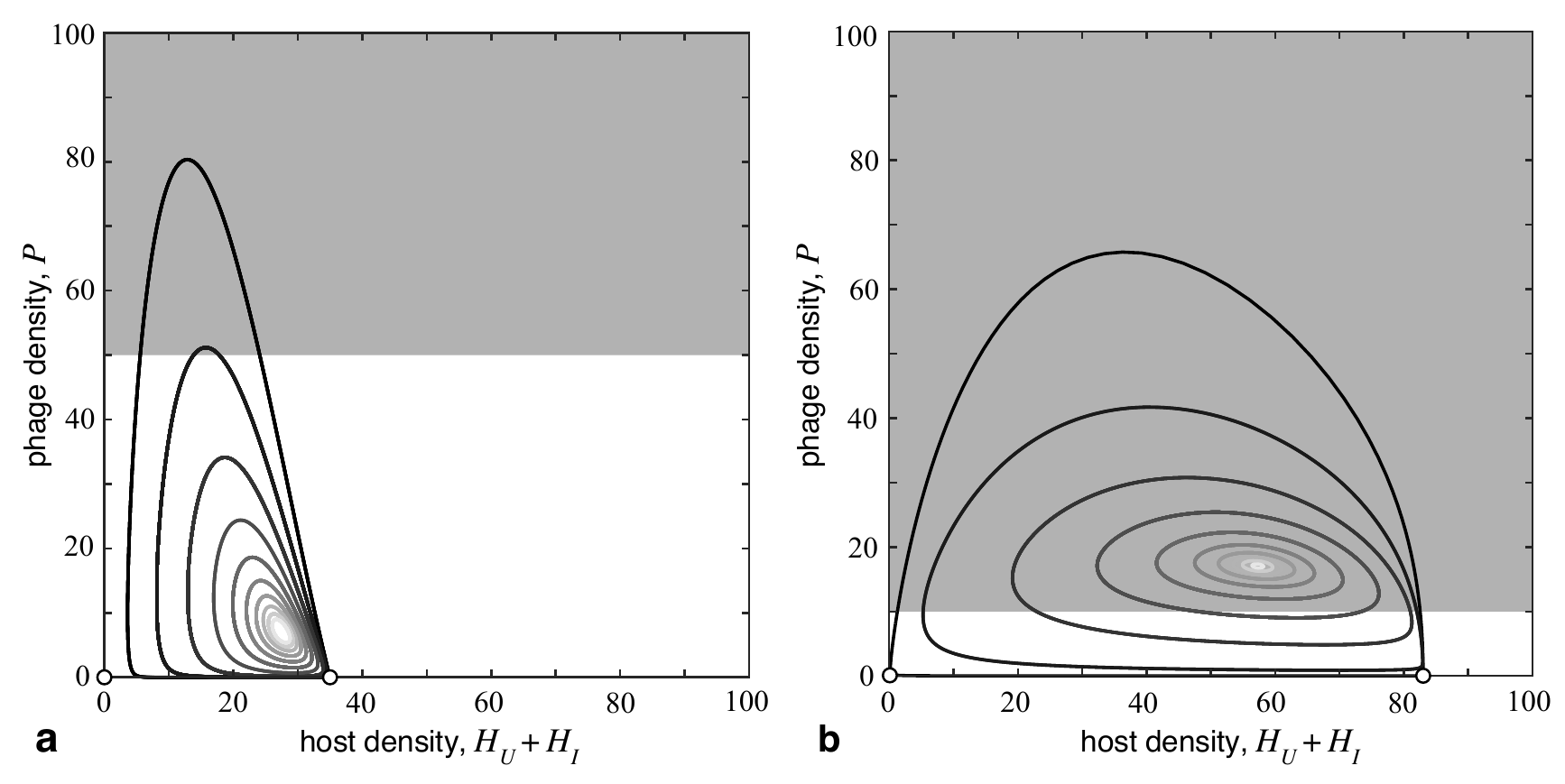}
\caption{Stable limit cycles for increasing $\lambda$: in 
\panelb{a} $r<\mu_1^2/\mu_2$ and $\lambda_H\approx 13.48$ while in 
\panelb{b} $r>\mu_1^2/\mu_2$ and $\lambda_H\approx 0.8328$. For $\lambda>\lambda_H$ a small limit cycle appears, which gradually grows with $\lambda$ (light to dark shades of grey). The shaded area marks regions for which phage densities are sufficiently high such that double infections dominate. Along the horizontal axis the open circles mark the locations of the two unstable equilibria: extinction, $E_0$, as well as the uninfected equilibrium, $E_U$. Note that the location of the interior equilibrium shifts with $\lambda$ (not shown, c.f. \figs{rl1}{rg1}). 
Parameters: $\xi=0.1,\mu_1=0.1,\mu_2 = 0.01, d=8.3,\kappa=1$ with 
\panelb{a} $r=0.7$, $\lambda\in\{13.50, 13.51, 13.54, 13.59, 13.67, 13.80, 14.05, 14.47, 15.20, 16.48\}$, and 
\panelb{b} $r= 8.3$, $\lambda\in\{0.8331,0.8335,0.8345,0.8368,0.8427,0.8577,0.8954,0.9901,1.2298,1.8329\}$.}
\label{fig:cycles}
\end{figure}

In both dynamical regimes, the size of the limit cycles for $\lambda>\lambda_H$ increases with $\lambda$, see \fig{cycles}. For larger $\lambda$, the limit cycles periodically alternate between the two domains of dominant single and double infections. In addition, as $\lambda$ increases the limit cycles approach vanishing phage densities and get closer to both the extinction equilibrium, $E_0$, as well as the uninfected equilibrium, $E_U$. As a consequence the dynamics become increasingly prone to stochastic effects, which can easily result in the extinction of the phage population or of both hosts and phages. 

The parameter $\kappa$, which sets the relative lifespan of phages versus hosts and thus impacts the density of phages in the environment and accordingly the rate of infections, has surprisingly little impact on the dynamics, see \app{kappa}.

\section{Discussion}
\label{section:Discussion}

Our model describes a phage-host system where multiple phages can simultaneously infect a single host. Considering the high densities of phage proposed for use in various applications, as well as the high densities of phage in many natural settings, simultaneous infections are a natural and relevant infection dynamic to consider. Our results shed light on several ecological features of this system and suggest interesting evolutionary implications.

\subsection{Impact of net bacterial growth rate}

We find that the qualitative dynamics depend on the net bacterial growth rate, $r$, compared to the ratio of the rates of single versus simultaneous double infections, $\mu_1^2/\mu_2$. An intuitive assumption posits that if single infections occur at a rate proportional to $\mu_1 P$ then it is natural that double infections occur proportional to $(\mu_1 P)^2$, which sets the threshold between the two dynamical regimes to $r=1$. Qualitatively different dynamics  occur if the rate at which a newly emerged uninfected host is doubly infected is higher versus lower than the rate at which simultaneously two separate uninfected hosts are singly infected.  

If bacterial growth rates are low ($r\leq\mu_1^2/\mu_2$), then $R_0<1$ is sufficient to eliminate phage infection, see \fig{rl1}, and the resulting dynamics are in excellent agreement with a previous study that models only single infection \citep{beretta:MB:1998}. Fittingly, in our model single infections always occur at a higher rate than double infections for low bacterial growth rates. Accordingly, our model can be interpreted as a generalized model that encompasses this previous work while further allowing for the investigation of  dynamical features arising through simultaneous infections. Indeed, for high bacterial growth rates  ($r>\mu_1^2/\mu_2$) dynamics are bi-stable and $R_0 <1 $ is necessary but not sufficient for phage elimination.  Bi-stability has important ramifications for both phage applications and disease management, which are discussed in full detail below. Interestingly, our results are in line with previous work on cooperative hunting in predator-prey systems, which similarly found that bi-stability in the predator population is only possible for sufficiently high prey growth rates \citep{alves:JTheorBiol:2017}.

The parameter $\kappa$ sets the relative life spans of hosts versus phages and thus influences the environmental density of phages. When bacterial growth rates are low ($r<\mu_1^2/\mu_2$), the dynamics are qualitatively identical regardless of $\kappa$. In contrast, when bacterial growth rates are high ($r>\mu_1^2/\mu_2$) the relative lifespans matter: if infected hosts outlive phages, ($\kappa\geq 1$) again no qualitative changes. However, if phages degrade more slowly than infected hosts lyse ($\kappa<1$), the dynamics admit a region where the infected equilibrium $E_{I^s}$ is a global attractor. Thus, for a range of burst sizes, $\lambda$, phages can invade and reach a stable density. Otherwise such dynamics are only observed for low bacterial growth rates.

\subsection{Limit cycles, extinction risk, and evolutionary implications} 

Phage-host dynamics often exhibit limit cycles, as is typical of predator-prey relationships. Specifically, in our model large $\lambda$ invariably result in stable limit cycles regardless of the host growth rate $r$: if uninfected hosts abound, phages always succeed in infecting a host but then large burst sizes result in a rapid increase in phage density, and phages quickly spread through the host population. As a consequence, the density of uninfected, susceptible hosts declines and phage replication slows down. The resulting decline of phage densities allows the hosts to recover and the cycle restarts. Experimental work with phages often requires controlling the relative density of phages and hosts, for example to control the expected number of bacteria that are infected by multiple phages \citep{turner:Nature:1999,turner:AmNat:2003}. Thus, even though controlling phage density in experiments may be inevitable, it may also obscure important ecological dynamics.

Further, the size of the stable limit cycles increase with $\lambda$ such that trajectories periodically get arbitrarily close not only to vanishing phage densities but also to the extinction equilibrium, see \fig{cycles}. In populations with small numbers stochastic effects matter \citep{traulsen:PRE:2012,huang:PNAS:2015}, and the smallest perturbation could easily result in the extinction of phages, or both phages and their host. In cases where phages drive the host population extinct they inevitably follow suit because they have exhausted their resource for replication. Thus while increasing the burst size $\lambda$  increases $R_0=r\mu_1\lambda/(d\kappa\xi)$ and hence makes it easier for phages to invade, larger $\lambda$  counter-intuitively also carry the inherent risk of extinction of phages or even both populations.  

Our work focuses on the \emph{ecological} impact of simultaneous infections but suggests an interesting \emph{evolutionary} question. In our model, host death is inevitable after infection. However, the burst size $\lambda$ influences phage density and accordingly the rate at which hosts are infected and then lyse. Consequently $\lambda$ makes a suitable proxy for virulence, or the increase in host mortality due to infection, as noted in the discussion of \cite{dennehy:PRSB:2004}. Though $\lambda$ is a key parameter in our analysis, it does not evolve. 

A natural expectation is that larger burst sizes would benefit the phage population. However, our analysis suggests that thereby phages also deplete their resources for further replication. Still, since selection acts first and foremost on the level of individuals, traits which produce more offspring should generally increase in frequency. This suggests tensions between selective pressures on individual phages and the viability of the phage population overall. In extreme cases this mismatch in individual fitness and population viability can lead to \emph{evolutionary suicide}, where viable populations adapt in a way that they can no longer persist \citep{parvinen:AB:2005,ferriere:PTRSB:2013}. Conversely, because of the interplay between evolutionary and ecological processes (such as fluctuation in host densities), parasites should evolve toward decreased virulence \citep{lenski:JTB:1994}. Moreover, co-infection of hosts (singly infected cells remain susceptible and can be infected again) \citep{alizon:RSIF:2013} can result in feedback loops between the rate at which susceptible individuals acquire a disease (the force of infection) and virulence: when double infections are common increased virulence is favored; however, increased pathogen virulence decreases the force of infection, favoring decreased virulence once more \citep{van-baalen:AmNat:1995}.

\subsection{Disease management and phage applications} 

When bacterial growth rates are high ($r>\mu_1^2/\mu_2$) simultaneous double infections result in bi-stabilities between the uninfected equilibrium, $E_U$, and stable co-existence of phages and hosts either at an interior equilibrium, $E_{I^s}$, or through periodic oscillations of a stable limit cycle. In particular, the bi-stability enables phages to \emph{persist} even though they are not able to \emph{invade} because their basic reproductive number $R_0$ is less than 1, see \fig{rg1}. This implies that a shock in phage densities would allow phages to get a foothold and persist.

Bi-stabilities have important ramifications for disease management \citep{dushoff:JMB:1998}. When bacterial growth rates are higher ($r>\mu_1^2/\mu_2$), ensuring $R_0<1$ is necessary but not sufficient to eliminate phages, see \fig{rg1}. Instead, it is required to either lower $R_0$ to a much greater extent, or to manipulate the ecology by introducing measures to lower phage densities through other means, which could make eliminating a problematic phage population difficult. However, it is enough to apply control measures just once to eliminate phages. Further, in the case of stable limit cycles, if timed right, a very small intervention may be sufficient to push the dynamics into the other basin of attraction, see \fig{pprl1} \& \ref{fig:pprg1}. In other words, there is the opportunity eliminate a detrimental strain with limited effort. 

 For high bacterial growth rates ($r > \mu_1^2/\mu_2$), in the region of bi-stability sufficiently densities of phages can persist even when they cannot invade. In the context of phage applications, introducing high densities of phage to a target bacterial population should be enough to ensure that the phage population takes off, even if the phage would not be able to establish from very low densities. Indeed, in the region of limit cycles, the necessary shock of phage density may be relatively low, implying that it may be possible to introduce a desired phage strain with limited effort.

\section{Acknowledgements}
We thank Alun Lloyd and Laura Parfrey for helpful discussion on an earlier version of the manuscript, as well as Arne Traulsen and Gabriel Currier for insightful comments. The authors acknowledge funding by the Natural Science and Engineering Research Council of Canada, Discovery Grant RGPIN-2021-02608 to C.H.

\section{Declaration of interest}
Declarations of interest: none

\bibliographystyle{abbrvnat}
\bibliography{ET_Dec2}

\newpage
\setcounter{page}{1}

\section*{Supplementary Material for The Impact of Simultaneous Infections on Phage-Host Ecology}
\subsection*{Jaye Sudweeks \& Christoph Hauert}
\supplement

\section{Full model derivation}
\label{app:FullModel}

In order to build our model, we consider microscopic reaction processes between phages and their bacterial hosts. First, in the absence of phages, the uninfected bacteria, $H_U$, divide at a rate $b_U$, perish at a rate $d_U$ and compete for resources with other hosts at rate $\xi$. The subscript $U$ refers to the fact that the hosts under consideration are uninfected. In terms of reaction kinetics this yields the following three equations:
\begin{subequations}
\label{eq:uninfectedHosts}
    \begin{alignat}{2}
        H_U &\reactto{b_U} &&\ H_U +H_U \\
        H_U &\reactto{d_U} &&\ \emptyset \\
        H_U + H_U &\reactto{\xi} &&\ H_U.
    \end{alignat}
\end{subequations}
Based on the mass-action principle, this results in classical logistic growth dynamics for uninfected hosts:
\begin{align}
\label{eq:logistic}
\dot{H}_U =&\ r\, H_U - \xi H_U^2,
\end{align}
where $r = b_U - d_U$ denotes the net per capita rate of growth. Naturally, \eq{logistic} admits two equilibria, the trivial equilibrium $H_0^* = 0$ and carrying capacity $H_U^*=r/\xi$. For $r<0$ the trivial equilibrium $H_0^*$ is stable and $H_U^*$ does not exist, or in other words is biologically not relevant, whereas for $r>0$, $H_0^*$ is unstable and $H_U^*$ is a stable equilibrium.

Similarly, the microscopic interactions between phages and bacteria can be captured by additional reaction kinetics equations. More specifically, we consider free living phages, $P$, that can infect hosts, and assume that infected hosts become passive in the sense that they are no longer able to reproduce or compete for resources with other hosts and perish generally at a faster rate than their uninfected counter parts, $d>d_U$. 

Single phages infect uninfected hosts at rate $\mu_1$, producing singly infected hosts, $H_{1}$, while two phages simultaneously infect an uninfected host cell at rate $\mu_2$, producing doubly infected host cells, $H_{2}$. Singly infected host cells can be infected again (at rate $\mu_1$) and become doubly infected hosts. We assume that host cells are never infected by more than two phages to ensure that the dynamics remain analytically tractable. This assumption  is supported by empirical evidence that infections are limited to few $\varphi_6$ phages per cell \citep{turner:JV:1999}; besides, very high phage densities are required for simultaneous triple infections to significantly affect the dynamics. 

After infection, phages replicate inside the host cell.  Once the phage particle count inside the host cell has reached a threshold, say $\lambda$ on average, the host lyses and releases phages back into the environment. We assume that lysing occurs once host resources are exhausted, and that the replication rate of phages is high, so the lysis time and final count of phage particles released is independent of whether one or two phages infected the host cell. Finally, we also assume that the phage densities are sufficiently higher than bacteria such that impacts of infection events on the phage pool can be neglected, see \app{phageDense}. The rate at which infected hosts die is $d$, while phages degrade at the rate $d\kappa$. Hence, for $\kappa>1$ phages are shorter lived than infected hosts, while for $\kappa<1$ phages outlive hosts. 

The corresponding reaction kinetics equations are:
\begin{subequations}
\label{eq:phageInfection}
    \begin{alignat}{2}
	H_U + P &\reactto{\mu_1} &&\  H_{1} + P\\
        H_{1} + P &\reactto{\mu_1} &&\  H_{2} + P \\
	H_U + P + P &\reactto{\mu_2} &&\  H_{2} + P + P \\
	H_{1} &\reactto{d} &&\ \lambda P \\
        H_{2} &\reactto{d} &&\ \lambda P \\
	P &\reactto{d\kappa} &&\ \emptyset 
    \end{alignat}
\end{subequations}

Note that single infection events happen at rate $\mu_1\,P$ while double infections happen at rate $\mu_2\,P^2$. Consequently it seems natural that each infection in a double infection event occurs at rate $\mu_1\,P$, and hence $\mu_2\,P^2 = (\mu_1\,P)^2$ such that $\mu_2 = \mu_1^2$. This allows us to eliminate one parameter by setting $\mu_1=\mu$ and $\mu_2=\mu^2$, but keeping the rates separate  provides further insights into interesting dynamical details. 

Following the mass-action principle, \eqtwo{uninfectedHosts}{phageInfection} yield the following system of dynamical equations for the densities of uninfected hosts, $H_U$, single infected hosts $H_{1}$, double infected hosts $H_{2}$, and free phage particles, $P$:

\begin{subequations}
\label{eq:intermediateEqns}
\begin{align}
    \dot{H}_U &= r\,H_U - \xi\,H_U^2 - \mu_1\,H_U\,P - \mu_2\,H_U\,P^2  \\
    \dot{H}_{1} &= \mu_1\,H_U\,P - \mu_1\,H_{1}\,P - d\,H_{1}  \\
    \dot{H}_{2} &= \mu_1\,H_{1}\,P + \mu_2\,H_U\,P^2 - d\,H_{2}  \\
    \dot{P} &= d\,\lambda\,(H_{1} + H_{2}) - d\,\kappa\, P .
\end{align}
\end{subequations}

Finally, and again for analytical tractability, we define $H_I =  H_{1} +  H_{2}$ and simplify our equations into  the following, final system of dynamical equations for the densities of uninfected hosts, $H_U$, infected hosts $H_I$ and free phage particles, $P$:
\begin{subequations}
\label{eq:appModelEqs}
\begin{align}
    \dot{H}_U &= r\,H_U - \xi\,H_U^2 - \mu_1\,H_U\,P - \mu_2\,H_U\,P^2 \label{eq:appdotHu} \\
    \dot{H}_I &= \mu_1\,H_U\,P + \mu_2\,H_U\,P^2 - d\,H_I \label{eq:appdotHi} \\
    \dot{P} &= d\,\lambda\, H_I - d\,\kappa\, P \label{eq:appdotP}.
\end{align}
\end{subequations}
Note that the term related to the conversion of single infected to double infected hosts was lost in the grouping to produce equations \eqtwo{appdotHu}{appdotP}. See \app{susWin} for an alternative derivation of our model that considers sequential infections only and assumes that a second infection can only occur in a small time window after the first infection. 

\section{Double infections through susceptible time window}
\label{app:susWin}
As an alternate derivation of our dynamical system, we can assume that only single infections occur (at rate $\mu_1$) but that a second, sequential infection can occur. However, we assume that singly infected hosts remain susceptible only for a short period, which is too brief for lysing to occur. This results in the following system of differential equations with $H_{1_S}$ denoting singly infected but susceptible hosts:

\begin{subequations}
\label{eq:SW_firstmodelEqs}
\begin{align}
    \dot{H}_U &= r\,H_U - \xi\,H_U^2 - \mu_1\,H_U\,P   \\
    \dot{H}_{1_S} &= \mu_1\,H_U\,P - \mu_1\,H_{1_S}\,P - \eta\,H_{1_S} \label{eq:1s}  \\
    \dot{H}_{1} &= \eta\,H_{1_S}  - d\,H_1  \\
    \dot{H}_{2} &= \mu_1\,H_{1_S}P  - d\,H_2  \\
    \dot{P} &= d\,\lambda\, (H_1 + H_2 ) -d\,\kappa\, P.
\end{align}
\end{subequations}

We assume a small susceptible time window, $\eta\gg\mu_1 P$. Further, $H_U\gg H_{1_S}$ because $H_{1_S} \ll H_I = H_1 + H_2$ and $H_I<H_U$ (recall our assumption that infected hosts die faster than uninfected hosts). Overall, hosts pass quickly through the susceptible state,  and thus the equilibration of susceptible infected hosts is reasonable. With the assumption $\eta\gg\mu_1 P$ we obtain at equilibrium $H^*_{1_S} \approx \frac{\mu_1 H_U P}{\eta}$.  This allows to eliminate \eq{1s} and insert the equilibrium to obtain:

\begin{subequations}
\label{eq:SW_finalmodelEqs}
\begin{align}
    \dot{H}_U &= r\,H_U - \xi\,H_U^2 - \mu_1\,H_U\,P - \mu_2\,H_U\,P^2 \\
    \dot{H}_{I} &= \mu_1H_{U}P + \frac{\mu_2}{\eta}H_{U}P^2 - dH_I \\
    \dot{P} &= d\lambda H_I -d\kappa P.
\end{align}
\end{subequations}

with $H_I=H_1+H_2$, which is equivalent to eq. 5 with $\mu_2=\mu_1^2/\eta$.

\section{Infection and phage density}
\label{app:phageDense}
Keeping phage densities constant in our model is conducive to analytical tractability and does not impact the qualitative dynamics. If infections deplete phage densities, then the reaction kinetic equations describing interactions between uninfected hosts and phages are
\begin{subequations}
\label{eq:denseAssumption}
    \begin{alignat}{2}
        H_U + P &\reactto{\mu_1} &&\ H_I \\
        H_U + P + P  &\reactto{\mu_2} && H_I.
    \end{alignat}
\end{subequations}
 The resulting  system of dynamical equations is identical to \eq{modelEqs} except the equation for $\dot{P}$, which becomes: 
\begin{equation}
\label{eq:VD_modelEqs}
    \dot{P} = d\,\lambda\, H_I - \mu_1\,H_U\,P - \mu_2\,H_U\,P^2  -d\,\kappa\, P.
\end{equation}

\begin{figure}[h!]
    \centering
    \includegraphics[width=0.9\linewidth]{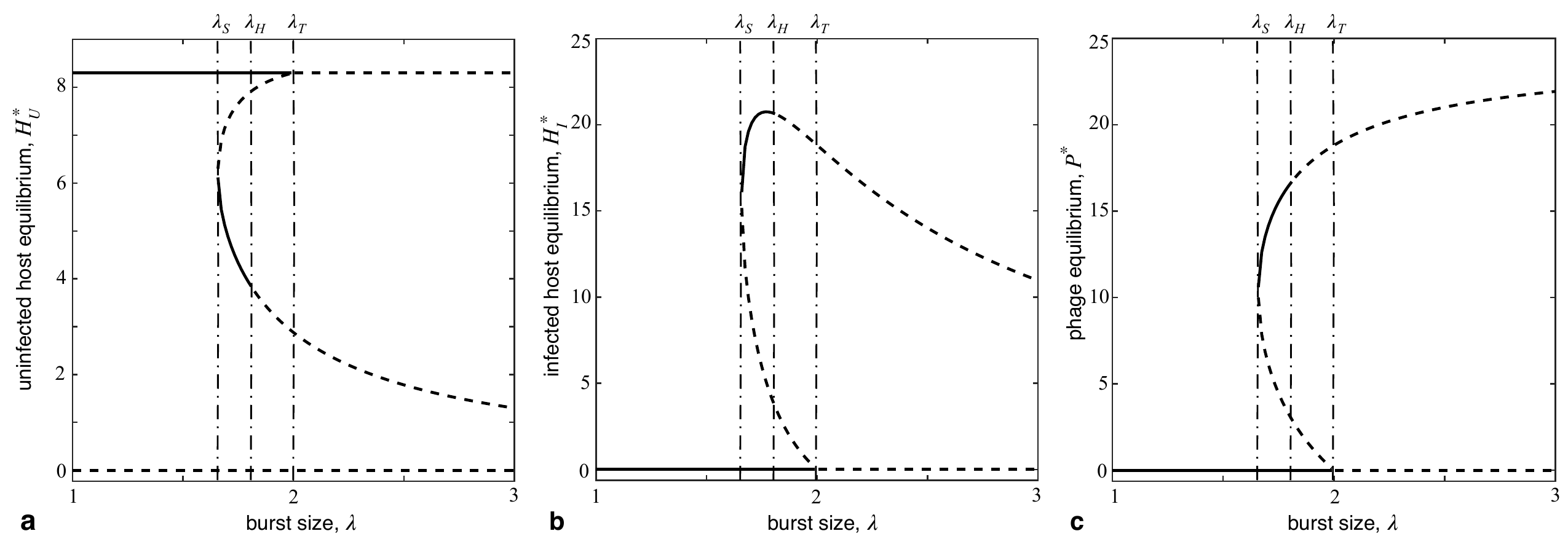}
    \caption{Bifurcation diagram for \eq{VD_modelEqs}. There are three bifurcation: first a saddle node bifurcation at $\lambda_S \approx 1.66$, next a Hopf bifurcation at $\lambda_H \approx 1.82$, and finally at transcritical bifurcation at $\lambda_T \approx 2.01$. This is the same series of bifurcations seen in the original model, though the exact values of the bifurcations have changed. All parameter values are identical to the manuscript ($r=8.3,d=8.3,\kappa=1,\mu_1 = 0.1, \mu_2 = 0.01, \xi = 0.1$).}
    \label{fig:VD_bifurcation}
\end{figure}

The resulting  bifurcation diagram exhibits the same qualitative dynamics as the original model, compare \fig{VD_bifurcation} to \fig{rg1} in the main text.

 We can re-derive $R_0$ as follows: To calculate the expected lifespan of a phage, or the expected length of time for which a phage has the potential to infect a host cell,  we must consider two terms: (i) $\mu_1 H_U P$  and (ii) $d\kappa P$.  Thus, when a phage is introduced into an entirely uninfected host cell population with $H_U = r/\xi$, the expected lifespan of a phage is $ {(\kappa d + \mu_1(r/\xi))}^{-1}$. The per capita rate of infection is $\mu_1$, so a phage introduced into an entirely uninfected host cell population infects $\mu_1\left( {r}/{\xi}\right)\left({(\kappa d + \mu_1(r/\xi)}\right)^{-1} $ cells on average. Thus, initially each phage produces $R_0= {\mu_1r\lambda}/{(\mu_1r + d\kappa\xi)}$ new phages, and phages can invade if $\lambda>1 + ({d\kappa\xi}/{\mu_1r})$.

\section{\label{app:jacobian}Stability analysis}
Because hosts undergo logistic growth in the absence of phages, we know the extinction equilibrium $E_0$ is unstable as long as $r>0$. The stability of other equilibria requires more work. After noting relationships between dynamical variables at equilibrium, we can write the Jacobian of the dynamical system, Eq. (1) in the main text, as:
\begin{align}
J =& \left(
\begin{array}{ccc}
 -\xi H_U  & 0 & -2d \frac{\kappa}{\lambda} + \mu_1 H_U \\
  r - \xi H_U& -d & 2d \frac{\kappa}{\lambda} - \mu_1 H_U \\
 0 & d \lambda  & -d \kappa  \\
\end{array}
\right).
\end{align}
An equilibrium is stable if the (real parts of) all eigenvalues are negative and unstable if at least one is positive. At the uninfected equilibrium, $E_U$, the Jacobian is
\begin{align}
\left(
\begin{array}{ccc}
 -r & 0 & \frac{-\mu _1 r}{\xi} \\
 0 & -d & \frac{\mu _1 r}{\xi } \\
 0 & d \lambda  & -d \kappa  \\
\end{array}
\right)
\end{align}
with eigenvalues $\{-r,-\Big[d(\kappa + 1) \pm \sqrt{\big(4 r d \lambda \mu_1 + d^2 \xi (\kappa-1)^2\big)/\xi}\,\Big]/2\}$, which are all negative for $\lambda<d\kappa\xi/(r \mu_1)$. In order to see this, consider the inequality
\begin{align}
\label{eq:negeig}
d(\kappa + 1) > \sqrt{\frac{4 r d \lambda \mu_1 + d^2 \xi (\kappa-1)^2}{\xi}}.
\end{align}
Note that both sides are positive. After squaring and rearrangements, we obtain
\begin{align}
\xi d^2 \big((\kappa + 1)^2 - (\kappa-1)^2 \big) = \xi d^2 4 \kappa > 4 r d \lambda\mu_1,
\end{align}
which yields the above threshold for $\lambda$.
\hfill\ensuremath{\square}

\section{Interior equilibria}
\label{app:interior}
\subsection{Derivation of biologically relevant interior equilibria}
\label{biorelIE}
The number of biologically relevant can be determined analytically based on the discriminant as well as Descartes’ rule of signs without having to deal with the unwieldy solutions to cubic equations.

The discriminant of a cubic polynomial of the form $p(x) = Ax^3 + Bx^2 + Cx + D$ is given by
\begin{align}
    \Delta = \frac{4\left(B^2-3AC\right)^3 - \left(2B^3 - 9ABC+27A^2D\right)^2}{27A^2}.
\end{align}
For $\Delta>0$, the polynomial $p(x)$ has three distinct real roots, for $\Delta=0$, either a triple root or two distinct real roots (one is a double root)  and finally for $\Delta<0$, only one real root remains (plus two imaginary roots, which are of no interest in the present context). 

In our case, $A = \lambda^4\mu_2^2$, $B = 2\kappa\lambda^3\mu_1\mu_2 $, $C = \kappa^2\lambda^2\left(\mu_1^2 - \mu_2\,r\right)$, and $D = {-\kappa^3\lambda\mu_1\,r + d\kappa^4\xi}$, c.f. Eq. (7) in the main text. After simplification, we find that $\Delta$ is proportional to a function which is quadratic in $\lambda$,
\begin{align}
\label{eq:deltan}
\Delta &\propto f(\lambda) := \mu_2\,r^2\left( \mu_1^2+4\,\mu_2 r \right) \lambda^2+ 2\,\,d\kappa\,\mu_1\xi \left( 2\,\mu_1^2\xi+9\,\mu_2\,r \right) \lambda-27\,d^2\kappa^2\mu_2\,\xi^2.
\end{align}
The function $f(\lambda)$ is an upwards facing parabola with two real roots:
\begin{align}
\label{eq:lambdapm}
\lambda_\pm := \frac{d\xi\kappa}{\mu_2r^2}\frac{-2\mu_1^3 - 9\mu_1\mu_2 r \pm 2(\mu_1^2 + 3\mu_2r)^\frac{3}{2}}{\mu_1^2+4\mu_2r}.
\end{align}
Clearly, the root $\lambda_-$, is always negative and thus not biologically relevant. However, $\lambda_+$ turns out to be always positive because $2(\mu_1^2 + 3\mu_2r)^{3/2}>\mu_1(2\mu_1^2 + 9\mu_2 r)>0$, which is easy to see after squaring both sides. Thus, whether $\lambda$ is smaller, equal to or larger than $\lambda_+$ determines the sign of $\Delta$ and accordingly the number of real roots. Hereafter, we refer to $\lambda_+$ as $\lambda_S$.

The number of positive roots can then be determined by Descartes' rule of signs, which states that the number of positive roots is  equal to the number of sign changes between consecutive, non-zero coefficients or an even number less than that. In Eq. (7) of the main text, $A$ and $B$ are always positive and $D$ is positive for $\lambda<d\kappa\xi/(r\mu_1)=:\lambda_T$. Note that $\lambda_S<\lambda_T$ always holds, see \app{lamSlamT}. The sign of $C$ depends on the growth rate $r$, which creates two general cases for the number of positive roots:
\begin{enumerate}[label=\emph{(\roman*)}]
\itemsep0pt
\item $r<\mu_1^2/\mu_2$: for $\lambda<\lambda_T$ no positive root and a single root for $\lambda>\lambda_T$;
\item $r=\mu_1^2/\mu_2$: in this non-generic case $C=0$ and $\lambda_S=\lambda_T$. Hence, no positive root for $\lambda<\lambda_T$ and still a single root for $\lambda>\lambda_T$;
\item $r>\mu_1^2/\mu_2$: no root for $\lambda<\lambda_S$, two positive roots for $\lambda_S < \lambda < \lambda_T$, and a single root for $\lambda > \lambda_T$.
\end{enumerate}
\subsection{$\mathbf{\lambda_S \leq \lambda_T}$}
\label{app:lamSlamT}
In order to show that $\lambda_S \leq \lambda_T$, we need to establish 
\begin{align}
\label{eq:lSlesslT}
 -\frac{d\xi\kappa}{\mu_2r^2}\frac{\mu_1(2\mu_1^2 + 9\mu_2 r) - 2(\mu_1^2 + 3\mu_2r)^\frac{3}{2}}{\mu_1^2+4\mu_2r} &\leq \frac{\kappa d \xi}{\mu_1 r}
\end{align}
which simplifies to
\begin{align}
\label{eq:lslt}
\mu_1 (\mu_1^2 + 3\mu_2r)^\frac{3}{2} &\leq \mu_1^4 + 5r \mu_1^2\mu_2 + 2 r^2 \mu_2^2.
\end{align}
Now consider the function $f(r):= \mu_1^4 + 5r \mu_1^2\mu_2 + 2 r^2 \mu_2^2 - \mu_1 (\mu_1^2 + 3\mu_2r)^\frac{3}{2}$, i.e. the difference between the two sides of the inequality \eq{lslt}. For $r\geq0$, $f(r)$ has roots at $r=0$ and $r=\mu_1^2/\mu_2$. The slope at $r=0$ is positive and the root at $r=\mu_1^2/\mu_2$ is a local minimum. As a consequence $f(r)\geq 0$ holds for $r>0$ and hence the inequality \eq{lslt} is always satisfied with equality only for $r=\mu_1^2/\mu_2$ (as well as in the limit $r\to 0$). 
\hfill\ensuremath{\square}

\section{$H_U^*$ is positive when $H_I^*$ is positive}
\label{app:boundHI}

From Eq. 7 of the main text we see that $H_U^*$ can be negative if $H_I^*>\kappa\Big(-\mu_1+ \sqrt{\mu_1^2+4\mu_2r}\Big)/(2\mu_2\lambda)$. Recall that $H_I^*$ are the roots of the cubic polynomial $f(H_I)$, Eq. 8 in the main text. Here we show that $f(H_I)$ has no roots at or above the threshold, meaning  $H_I^*$ is bounded below the threshold, and consequently $H_U^*$ is always positive when $H_I^*$ is positive.

 We denote the threshold $\hat{H_I}:=\kappa\Big(-\mu_1+ \sqrt{\mu_1^2+4\mu_2r}\Big)/(2\mu_2\lambda)$. First note that  $f(\hat{H_I}) = {{d{\kappa}^{4}\xi}/{\lambda}} >0$. Therefore $\hat{H_I}$ is not a solution to $f(H_I)$, so $H_I^* \neq \hat{H_I}$. We now show that $f(H_I)>0$ for $H_I>\hat{H_I}$ and does not admit solutions to $f(H_I)=0$ for $H_I>\hat{H_I}$. 

First, $  f'(\hat{H_I}) = \frac{1}{2}{\lambda\,{\kappa}^{2} \left( \mu_1\,\sqrt {{\mu_1}^{2}+4\,\mu_2\,r}+
{\mu_1}^{2}+4\,\mu_2\,r \right)} >0$. Second, the second derivative of $f(H_I)$ is 
\begin{align*}
    f''(H_I) = 6\,H_I{\lambda}^{4}{\mu_2}^{2}+4\,\kappa\,{\lambda}^{3}\mu_1\,\mu_2,
\end{align*}

which is positive for $H_I>0$. Therefore $f'(H_I) > 0$ for $H_I>\hat{H_I}$. In summary, $f(H_I)$ is positive at $\hat{H_I}$ and only increases, and consequently no roots $H_I^* \geq \hat{H_I}$ exist.

\hfill\ensuremath{\square}

\section{Single infections dominate at $\mathbf{E_I}$ for $\mathbf{r<2\mu_1^2/\mu_2}$}
\label{app:VstarAnalysis}

At the interior equilibrium, $E_I$, single infections dominate for sufficiently small $r$. More specifically, single infections dominate when  $P^*<\mu_1/\mu_2$. Using \eq{Pstar} this condition becomes $H_I^*<\kappa\mu_1/(\lambda\mu_2)$. After setting $x = H_I^*\lambda\mu_2/(\kappa\mu_1)$ in \eq{cubicHi} of the main text, some algebra yields
\begin{align}
    \label{eq:Ceq}
    \frac{d \kappa \xi}{\lambda \mu_1}+(x+1) \left(\frac{\mu_1^2}{\mu_2} x (x+1)-r\right) = 0,
\end{align}
which at the threshold, $x = 1$, simplifies to
\begin{align}
\label{eq:Ce1}
	\frac{d \kappa \xi}{2\lambda \mu_1} + 2\frac{\mu_1^2}{\mu_2} - r = 0.
\end{align}
Thus, the left hand side (LHS) of \eq{Ce1} is strictly positive for $r<2\mu_1^2/\mu_2+d\kappa\xi/(2\lambda\mu_1)$, which means that no solution exists to \eq{Ceq} and hence $P^*=\mu_1/\mu_2$ never holds. Note that the only negative term of \eq{Ceq} involves $r$ and hence the LHS is minimized when $r$ attains its maximum value. Instead of the maximum we use a smaller $r=2\mu_1^2/\mu_2$ to simplify the analysis and \eq{Ceq} becomes
\begin{align}
    \label{eq:Ceqbigr}
	\frac{d \kappa \xi}{\lambda \mu_1}+\frac{\mu_1^2}{\mu_2} \left(x^2-1\right) (x+2) = 0.
\end{align}
No solutions to \eq{Ceqbigr} with large $x$ exist because the LHS is strictly positive. However, for $x<1$ one term becomes negative. For any solution to \eq{Ceqbigr}, $x<1$ must hold, and hence $P^*<\mu_1/\mu_2$ holds for $r\leq 2\mu_1^2/\mu_2$. Thus, for all $r$ in our first dynamical regime $r<\mu_1^2/\mu_2$ we know that single infections dominate at $E_I$.
\hfill\ensuremath{\square} \\

\vspace{1cm}
However, for $r>2\mu_1^2/\mu_2$,  solutions with $x = 1$, or equivalently $H_I^* = \kappa\mu_1/(\lambda\mu_2)$, must hold for sufficiently large $\lambda$. Solving \eq{Ce1} yields the $\hat{\lambda}$ for which $H_I^* =  \kappa\mu_1/(\lambda\mu_2)$:
\begin{align*}
    \hat{\lambda} &= \frac{d\kappa\xi\mu_2}{2\mu_1(\mu_2 r - 2\mu_1^2)}.
\end{align*}
Interpreting $ \hat{\lambda}$ requires closer consideration. When $r>\mu_1^2/\mu_2$ (bi-stability possible), a pair of equilibria branch from a single point at the saddle node bifurcation $\lambda_S$. We can find expressions for this point by finding where \eq{cubicHi} has a double root. For our analysis, we need only consider the $P$ component of the double root, which we'll denote $Z=g(r)$, as the location of the double root is dependent on $r$.

In the following analysis, we assume that $\mu_1 = \mu$ and $\mu_2 = \mu^2$. When $r=8$, then $Z = 1/\mu$ and is located on the boundary between the regions where single versus double infections dominate. This means that $\hat{\lambda}=\lambda_S$, and the top branch $E_{I_U}$ is always in the region where single infections dominate while the bottom branch $E_{I_S}$ is always in the region where double infections dominate. When $r>8$, then  $Z$ is located in the region where double infections dominate and $\hat{\lambda}$ corresponds to where the bottom  branch ($E_{I_U}$) crosses into the region where single infections dominate, compare to \fig{rg1}\panel{C}. Finally, when  $2(\mu_1^2/\mu_2)=2 < r < 8$, then $Z$ is located in the region where single infections dominate, but the top branch ($E_{I_U}$) will cross the threshold at $\hat{\lambda}$.    

It follows that when $r>2\mu_1^2/\mu_2$, either double infections always dominate at the stable interior equilibrium $E_{I^s}$, or else double infections will eventually dominate at that equilibrium  for sufficiently large $\lambda>\hat\lambda$.

\section{Parameters \& comparison to previous results}
\label{app:parameters}

The model of \citep{beretta:MB:1998}, hereafter B\&K, is similar in form to \eq{modelEqs} but considers only single infections. Accordingly, \eq{modelEqs} can be interpreted as a generalized model that encompasses B\&K but also reveals interesting dynamical features arising through simultaneous infections. As such, a comparison of results from B\&K and \eq{modelEqs} is of interest. This comparison is particularly relevant because B\&K use $r >\mu_1^2/\mu_2$, part of the parameter case where \eq{modelEqs} exhibits bi-stability. This indicates that simultaneous double infections impact dynamics for B\&K's parameters, an effect that their model cannot account for. 

\renewcommand{\arraystretch}{2} 
\begin{table}[h!]
\begin{tabular}{|c|c|c|}
\hline
Parameter & Value & Comparison to parameter values in \citep{beretta:MB:1998} \\ \hline
\multirow{2}{*}{r} & ($r> \mu_1^2/\mu_2$) : 8.3 & same order of magnitude as corresponding $\alpha$ \\ \cline{2-3} 
 & ($r<   \mu_1^2/\mu_2$) :  0.7 & one order of magnitude smaller than above \\ \hline
$\xi$ & 0.1 & differs from equivalent ($\alpha/C$) by $10^6$ \\ \hline
$\mu_1$& 0.1& differs from equivalent $K$ by $10^7$. \\ \hline
$\mu_2=\mu_1^2$ & 0.01 & no double infections in B\&K \\ \hline
d & 8.3 & same order of magnitude as corresponding $\lambda$ \\ \hline
$\kappa$ & 1 & \makecell{viral death rate $d\kappa$ on same order of magnitude \\ as corresponding parameter $\mu$}  \\ \hline
$\lambda$ & bifurcation parameter & corresponds to bifurcation parameter $b$   \\ \hline
\end{tabular}
\caption{}
\label{tab:params}
\end{table}

Column 1 of Table \ref{tab:params} lists our parameter values while column 2 compares our parameters to those from B\&K. Most analogous parameters are on the same order of magnitude, but two parameters are not closely aligned: $\xi$ differs from its equivalent ($\alpha/C$) by $10^6$ while $\mu_1$  differs from its equivalent $K$ by $10^7$. However, these differences can be understood as a rescaling of density, see \app{rescalingDensity} for more details. 

As such, we can compare and contrast our results for $r>\mu_1^2/\mu_2$ to those of B\&K. There are similarities between the results: for small $\lambda$ the uninfected equilibrium is the only biologically relevant equilibrium, for $R_0 >1$ phage can invade, and both models exhibit limit cycles for sufficiently high $\lambda$. These similarities support the interpretation of \eq{modelEqs} as a generalization of B\&K.
 
 However, there is a crucial difference: \eq{modelEqs} admits a region of bi-stability where phages introduced through a sufficiently large shock can co-exist with hosts, either at a stable internal equilibrium or through the oscillations of a limit cycle, though $R_0<1$. In other words, phages can persist even though they cannot invade. Indeed, in the regime of limit cycles small shocks in phage density are sufficient to establish stable co-existence despite $R_0<1$. In summary, accounting for simultaneous double infections predicts that there is a range of $\lambda$ where, though $R_0<1$, hosts are vulnerable to sustained infection after shocks in phage density. 
 
\subsection{Rescaling density}
\label{app:rescalingDensity}
The large differences between our $\xi$ and $\mu$ and the analogous parameters ($\alpha/C$)  and $K$ in B\&K can be understood as a rescaling of density. To see this, consider respective rescaled densities of uninfected hosts, infected hosts, and phages: $ X = (1/c_1)H_U$, $Y = (1/c_2)H_I$, $Z = (1/c_3)P$. Assuming $\mu_1=\mu$, $\mu_2 = \mu_1^2$, and that $c_1=c_2$, or that uninfected and infected hosts are rescaled in the same way, \eq{modelEqs} becomes 

\begin{subequations}
\label{eq:rescaleSimplified}
    \begin{align}
       \dot{X} &=  r\,X - \hat{\xi}\,X^2 - \hat{\mu}\,XZ - \hat{\mu}^2\,X\,Z^2  \\ 
       \dot{Y} &= \hat{\mu}\,X\,Z + \hat{\mu}^2\,X\,Z^2 - d\,Y  \\
       \dot{Z} &= \,d\hat{\lambda}\,Y - d\,\kappa\,Z
    \end{align}
\end{subequations}

 where $\hat{\xi} = c_1\xi$, $\hat{\mu} = c_3\mu$, and $\hat{\lambda} = (c_2/c_3)\lambda$.

In interpretation, \eq{rescaleSimplified} indicates that rescaling density results in a rescaling of three parameters: $\xi$, $\mu$, and $\lambda$. This is consistent with a comparison between our parameters and results and those of B\&K: all analogous parameters are on the same order of magnitude with the exception of $\xi$ and $\mu$, and bifurcation values have shifted, consistent with a rescaling of the bifurcation parameter $\lambda$.

\clearpage\newpage
\section{Longevity of hosts versus phages, $\kappa$}
\label{app:kappa}
For $\kappa<1$ phages are longer lived and their densities in the environment are higher. This increases infections such that lower burst sizes $\lambda$ are required to trigger qualitative changes in dynamics. The saddle node bifurcation $\lambda_S$, and transcritical bifurcation $\lambda_T$ are linear functions of $\kappa$ and hence maintain their ordering. However, because we have no analytical expression for $\lambda_H$, we turn to numerical analysis to determine the impact of $\kappa$.

\begin{figure}[tbp]
\centerline{\includegraphics[width=0.9\textwidth]{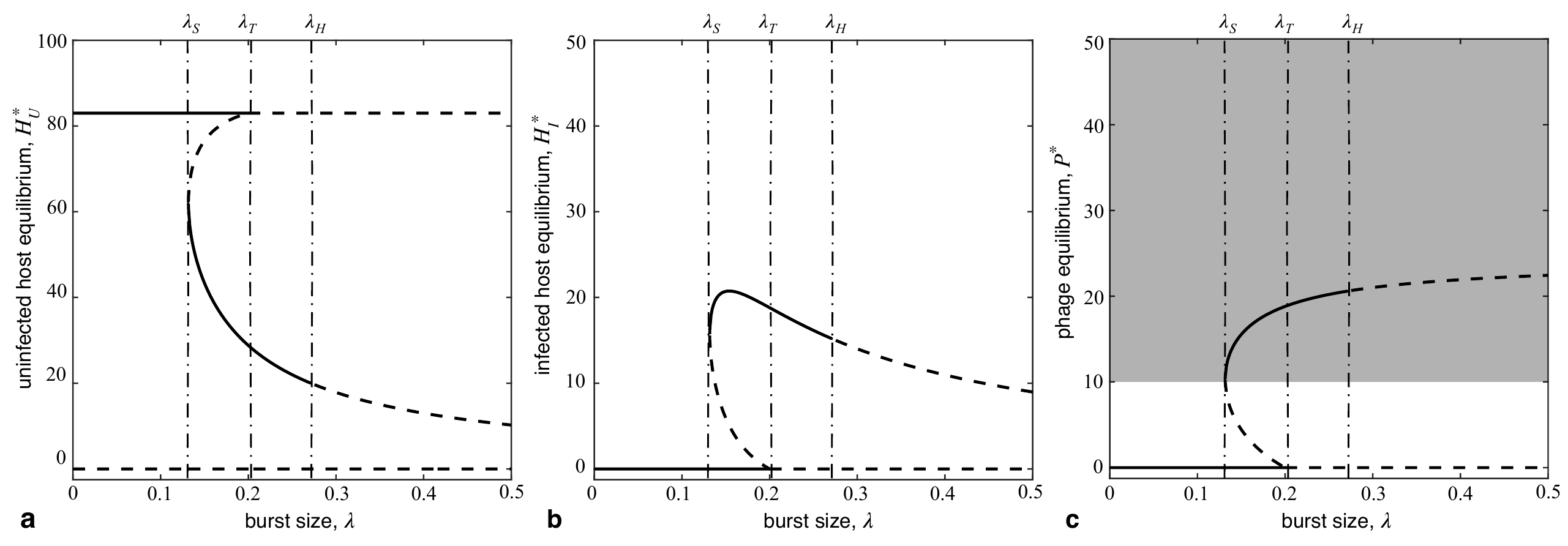}}
\caption{Bifurcation diagram for the dynamics in Eq. (4) of the main text with $\kappa<1$ and $r>\mu_1^2/\mu_2$ as a function of the burst size $\lambda$. Each panel depicts one component of the stable (solid lines) and unstable (dashed lines) equilibria, $E_i = (H_U^*, H_I^*, P^*)$: 
\panelb{a} uninfected hosts, $H_U^*$, 
\panelb{b} infected hosts, $H_I^*$, and 
\panelb{c} phage densities, $P^*$, where the shaded area marks densities for which double infections dominate. The trivial equilibrium, $E_0$, is always unstable and the uninfected equilibrium $E_U$ is stable for small $\lambda$. There are three bifurcations (dash-dotted vertical lines): a saddle node bifurcation at $\lambda_S = 0.1317$ (c.f. $\lambda_-$ in \eq{lambdapm}) where a pair (stable, $E_{I^s}$ and unstable, $E_{I^u}$) of interior equilibria appear; a transcritical bifurcation at $\lambda_T = 0.2000$ where the (originally) unstable interior equilibrium, $E_{I^u}$, disappears and $E_U$ loses stability; and finally Hopf-bifurcation at at $\lambda_H \approx  0.2707$ 
where $E_{I^s}$ loses its stability. This results in four dynamical regimes: for $\lambda<\lambda_S$ the uninfected equilibrium, $E_U$, is stable and a global attractor; for $\lambda_S<\lambda<\lambda_T$ the dynamics are bi-stable with attractors $E_U$ and $E_{I^s}$; for $\lambda_T<\lambda<\lambda_H$ the infected equilibrium $E_{I^s}$ is stable and a global attractor; finally, for $\lambda>\lambda_H$ no stable equilibria exist but there is a stable limit cycle, see \fig{ppkl1rg1} \panel{d}, \panel{h}. Parameters: $r=8.3,\xi=0.1,\mu_1=0.1,\mu_2 = 0.01, d=8.3,\kappa=0.2$ (same as in Fig. 3 except $\kappa$).}
\label{fig:kl1rg1}
\end{figure}

\begin{figure}[tbp]
\centering\includegraphics[width=0.9\linewidth]{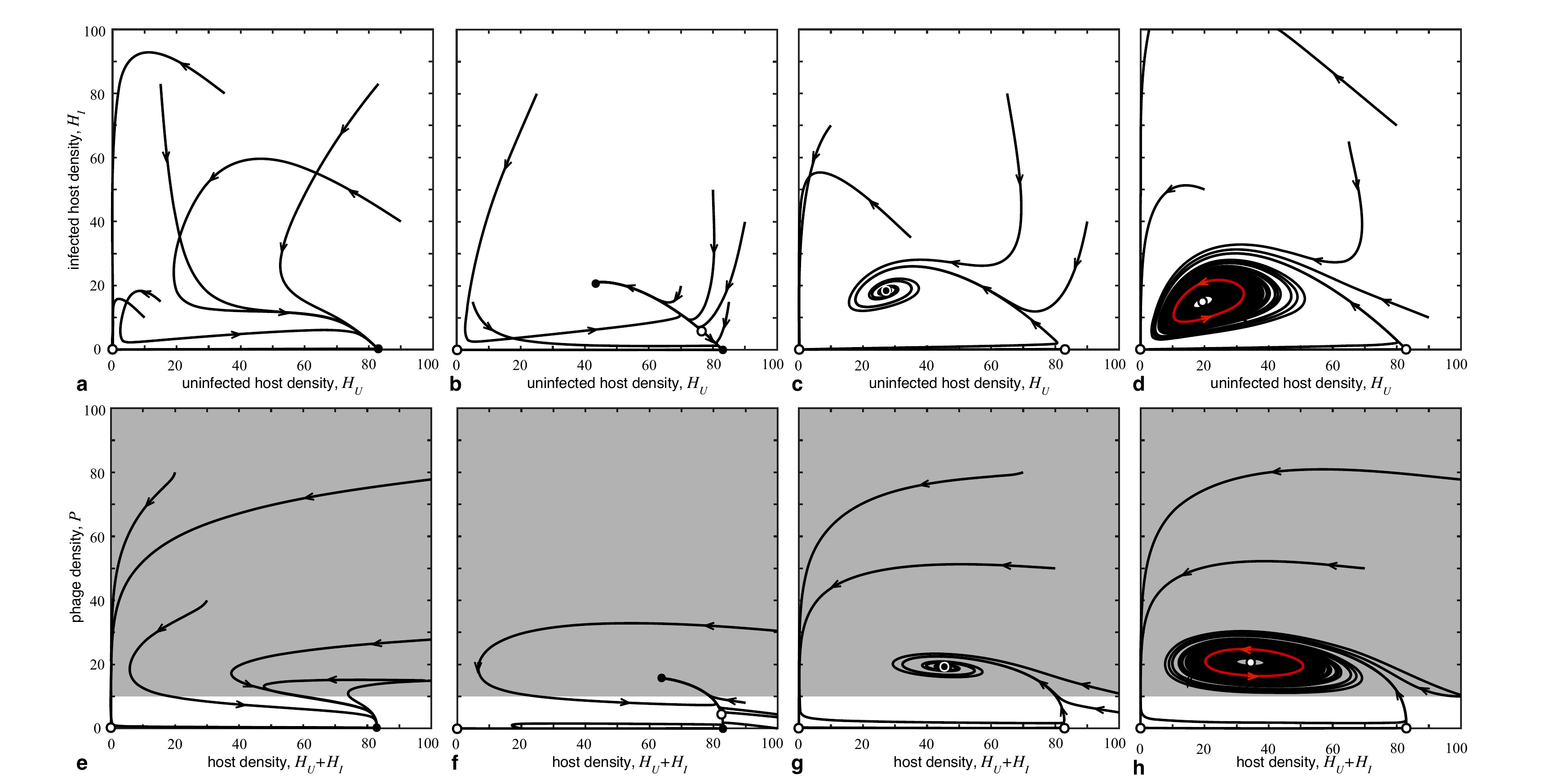}
\caption{Projections of the phase space depicting the characteristics of the four dynamical regions for $\kappa <1$ and $r>\mu_1^2/\mu_2$. The top row shows infected, $H_I$, versus uninfected, $H_U$, host densities and the bottom row shows phage densities, $P$, versus host densities, $H_U+H_I$. Shaded regions on the bottom row mark phage densities for which double infections dominate. Note that trajectories may cross because the panels show projections of a three dimensional phase space. Dots mark stable (filled) and unstable (open) equilibria. Stable limit cycles are shown in red.
\panelb{a, e} $\lambda = 0.05$: all trajectories converge to $E_U$ and phages go extinct.
\panelb{b, f} $\lambda = 0.15$: a pair of interior equilibria appear, $E_{I^u}$ and $E_{I^s}$, and the dynamics become bi-stable with attractors $E_U$ and $E_{I^s}$.
\panelb{c, g} $\lambda = 0.21$: the (originally) unstable interior equilibrium $E_{I^u}$ disappears, $E_U$ loses stability, and all trajectories go to the stable global attractor $E_{I^s}$. 
\panelb{d, h} $\lambda = 0.28$: $E_{I^s}$ loses stability and only a globally stable limit cycle remains.
Parameters: same as in \fig{kl1rg1}.}
\label{fig:ppkl1rg1}
\end{figure}

For $r< \mu_1^2/\mu_2$, the ordering of bifurcations remains the same $\lambda_S<\lambda_T<\lambda_H$ regardless of $\kappa$ and the dynamics remain qualitatively unchanged, see Figs. 1 \& 2 in the main text. Similarly, for $r>\mu_1^2/\mu_2$ and $\kappa\geq1$ the dynamics again remain qualitatively unchanged, see Figs 3 \& 4 in the main text, and the ordering of bifurcations remains $\lambda_S<\lambda_H<\lambda_T$. Only for $r>\mu_1^2/\mu_2$ and $\kappa<1$ the ordering changes to $\lambda_S<\lambda_T<\lambda_H$, see \fig{kl1rg1}.
Most importantly, the interior equilibrium becomes a global attractor for $\lambda_T<\lambda<\lambda_H$, see \fig{ppkl1rg1}\panel{c}, \panel{g}.
This dynamical behavior was seen previously for $r<\mu_1^2/\mu_2$ (see Fig. 1\panel{b}, \panel{e} in the main text). In both scenarios, the transcritical bifurcation $\lambda_T$ occurs before the Hopf bifurcation $\lambda_H$.

\end{document}